%% file: paper.tex
\documentclass[conference]{IEEEtran}
\IEEEoverridecommandlockouts
\usepackage{cite}
\usepackage{amsmath,amssymb,amsfonts}
\usepackage{algorithmic}
\usepackage{graphicx}
\usepackage{textcomp}
\usepackage{xcolor}
\usepackage{xspace}

\newcommand{\zephyr}{Zephyr\xspace}

\usepackage[firstpage=true,placement=bottom]{background}
\usepackage{tikz}
\newcommand{\faudisclaimer}{
  \begin{tikzpicture}[remember picture, overlay, black]
    \node{
      \begin{minipage}{\textwidth}
        \setlength{\fboxsep}{1em}%
        \newlength{\copylen}\setlength{\copylen}{\textwidth}\addtolength{\copylen}{-2\fboxsep}%
        \colorbox{yellow!70!black!20}{%
          \parbox{\copylen}{%
            \sffamily\scriptsize{}%
            \textcopyright{} 2021 IEEE. Personal use of this material is permitted. Permission from IEEE must be
            obtained for all other uses, in any current or future media, including
            reprinting/republishing this material for advertising or promotional purposes, creating new
            collective works, for resale or redistribution to servers or lists, or reuse of any copyrighted
            component of this work in other works.
            This is the author's version of the work. The definitive Version of Record was published in
            2021 XI Brazilian Symposium on Computing Systems Engineering (SBESC'21),
            November 22--26, 2021, Florianopolis, Brazil,
            https://doi.org/10.1109/SBESC53686.2021.9628358
          }}
        \vspace{4mm}
      \end{minipage}
    };
  \end{tikzpicture}
}
\SetBgContents{\faudisclaimer}
\SetBgOpacity{1.0}
\SetBgAngle{0.0}
\SetBgScale{1.0}

\begin{document}

\title{Migration-Based Synchronization
}

\newcommand{\affilfau}[1]{
 \IEEEauthorblockA{\textit{Friedrich-Alexander-Universit\"at Erlangen-N\"urnberg}\\
 \textit{#1}}
}
\newcommand{\affilrub}[1]{
 \IEEEauthorblockA{\textit{Ruhr University Bochum}\\
 \textit{#1}}
}

\author{%
 \IEEEauthorblockN{Stefan Reif, Phillip Raffeck, Luis Gerhorst, Wolfgang Schr\"oder-Preikschat}
 \affilfau{\{reif,raffeck,gerhorst,wosch\}@cs.fau.de}
 \and
 \IEEEauthorblockN{Timo H\"onig}
 \affilrub{timo.hoenig@rub.de}
}

\maketitle

\input{sec/abstract}


\input{sec/introduction}
\input{sec/background}
\input{sec/implementation}

\input{sec/evaluation}

\input{sec/conclusion}

\section*{Acknowledgments}
\footnotesize{
This work was partially funded by the Deutsche Forschungsgemeinschaft (DFG) — project number 465958100 (HO 6277/1-1 and SCHR 603/16-1) and by the Bundesministerium f\"ur Bildung und Forschung (BMBF) – project AI-NET-ANTILLAS (16KIS1315).
}

\bibliographystyle{IEEEtran}
\bibliography{bib/references}

\end{document}

%% file: sec/abstract.tex
\begin{abstract}
A fundamental challenge in multi- and many-core systems is the correct execution of concurrent access to shared data.
A common drawback from existing synchronization mechanisms is the loss of data locality as the shared data is transferred between the accessing cores.
In real-time systems, this is especially important as knowledge about data access times is crucial to establish bounds on execution times and guarantee the meeting of deadlines.

We propose in this paper a refinement of our previously sketched approach of Migration-Based Synchronization~(MBS) as well as its first practical implementation.
The core concept of MBS is the replacement of data migration with control-flow migration to achieve synchronized memory accesses with guaranteed data locality.
This leads to both shorter and more predictable execution times for critical sections.
As MBS can be used as a substitute for classical locks, it can be employed in legacy applications without code alterations.

We further examine how the gained data locality improves the results of worst-case timing analyses and results in tighter bounds on execution and response time.
We reason about the similarity of MBS to existing synchronization approaches and how it enables us to reuse existing analysis techniques.

Finally, we evaluate our prototype implementation, showing that MBS can exploit data locality with similar overheads as traditional locking mechanisms.

\end{abstract}

%% file: sec/introduction.tex
\section{Introduction}
\label{sec:introduction}

With ongoing improvements in chip production, it is only a matter of time until the many-core age also begins in the embedded domain.
Some existing platforms already head in this direction (6 core TriCore~\cite{infineon:20:tc39}, 8 core RISCV GAPuino~\cite{greenwaves:20:gap8}, 36 core ARM Marvell OCTEON~\cite{marvell:20:octeon}) and the usage of such platforms in systems, which are subject to timeliness constraints, is also only a question of time.
This development will lead to scenarios where also in embedded systems cores are available in abundance, meaning that not all cores are required to guarantee timely and efficient system execution.
We propose that this allows us to rethink common design concepts, in the case of this paper, thread synchronization.

The additional computing power that comes with additional cores is, however, not for free, but rather comes with the price of increasing architectural complexity.
Modern platforms come with a variety of memory hierarchies, leading to different access times from different cores.
This hierarchy ranges from fast core-local memories to equally accessible global memory regions.
Access to global memory is, however, often sped up via core-local caches with various coherency guarantees.

In general, it is difficult (up to impossible) to predict cache behavior, and more complex cache and memory hierarchies only aggravate the problem.
From a timing analysis perspective, this leads to overly pessimistic results, which negatively impact schedulability and system design.
Measurement-based analysis techniques mitigate the problem partly but only yield reliable results if the scenarios covered in the measurements are representative and complete.
Complex interferences between processors due to, for example, memory coherency, complicate such assessments of measurement quality.

In systems with completely independent threads or only a few inter-thread dependencies, these difficulties related to the memory hierarchy can potentially be avoided by partitioning the system into multiple, independent single-core systems.
With rising complexity of the applications, however, this approach becomes infeasible, inevitably leading to the sharing of data across multiple cores.

A variety of, both general-purpose and specialized, protocols exists to enable synchronization between different accesses to shared data and guarantee correct system behavior.
The employed techniques range from the use of locks over message passing to the wait-free design of algorithms.
Their usage further hinders the predictability of the system behavior, as analyses now also need to consider possible access patterns to shared resources and provide bounds for interferences.

Except for the special case of non-cached global memory accesses, sharing data between cores always entails migration of at least part of the shared data between cores.
This data migration is the source of the aforementioned interference cost, which is hard to predict because of the uncertainty which part of the data is already cached locally and which has to be migrated.
In contrast to the migration of application data, control-flow migration is highly predictable when the exact point of execution is known beforehand~\cite{klaus:19:ospert}, as the currently used thread-local data is not used by other threads and is determinable by static analysis.
For lock acquisition, those points are easily identifiable in the code.

When designing security-relevant or real-time systems, typical optimization goals are throughput and predictability.
Oftentimes, these goals are in conflict with each other.
A simple example for such a conflict is the use of caches for shared data: Temporarily storing shared data in core-local caches reduces access times and leads to a speed-up in execution time compared to storing it in non-cached global memory.
On the other hand, caching data shared between different cores complicates the prediction of access times (Will the access result in a hit or miss in the core-local cache?) and thus timing analyses, leading to results far more pessimistic than the actual executions.

In this paper, we sharpen our previously proposed concept of \emph{Migration-Based Synchronization~(MBS)}~\cite{reif:20:rtsswip} to increase the predictability of the system behavior.
As a side-effect, MBS can also lead to an increase in throughput, by  ensuring and exploiting data locality wherever possible.

The first step towards this goal is to exploit the abundance of cores to use some cores exclusively as central servers (i.e., ``synchronization cores'') for the operation with a single shared resource.
With such a strict separation, all threads sharing a resource benefit from data locality while accessing the shared resource, facilitating the analysis of such critical sections, as knowledge about the state of the core-local memory of the synchronization cores can be utilized.
Thus, in this paper, we propose to replace the migration of shared resources with the migration of control flow to obtain a more easily predictable system.
We strive to design MBS in such a way that it is transparently replaceable with traditional locks.
In this sense, MBS is the natural adaption of existing lock-based multi-core synchronization protocols to an abundance of cores in modern platforms.
On top, depending on the nature of the overall application and the usage pattern of shared resources, MBS is even able to outperform existing approaches.

In short, we provide the following contributions:
\begin{itemize}
  \item We present a refinement of our previously proposed, control-flow-based synchronization technique MBS.
  \item We provide the first implementation of that technique in the real-time operating system Zephyr.
  \item We evaluate the functionality and effectiveness of our prototype on real hardware.
  \item We discuss the influence of MBS on timing analysis supported by timing analysis experiments.
\end{itemize}

The rest of the paper is structured as follows.
Section~\ref{sec:background} presents an overview of the necessary background and related work.
The general concept of MBS and its effect on timing analysis as well as a prototype implementation are introduced in Section~\ref{sec:implementation}.
Section~\ref{sec:evaluation} presents experimental results evaluating the overheads of MBS and the potential improvements in timing analysis.
Section~\ref{sec:conclusion} concludes the paper.

%% file: sec/background.tex
\section{Background and Related Work}
\label{sec:background}

Synchronizing concurring threads through means of shared memory or message passing is a long researched topic~\cite{andrews:83:cs}.
As the interaction of threads over shared resources greatly influences their timing behavior, synchronization mechanisms are also a highly researched topic in the real-time community~(an overview can be found in~\cite{brandenburg:2019:arxiv}).
In particular, shared resources can lead to \emph{priority inversion}, meaning ready-to-run higher priority threads cannot execute because of blocked resources held by lower priority threads~\cite{brandenburg:13:rtas}.
In other words, priority inversions are any deviation from the expected schedule in contrast to normally expected interference by higher priority threads~\cite{yang:2015:rtss}.

Very similar to our proposed approach are the seminal multi-core versions of the priority-ceiling protocol by Rajkumar et al., which exist in a shared-memory~\cite{rajkumar:90:icdcs} and a message-passing~\cite{rajkumar:88:rtss} version, also referred to as MPCP and DPCP, respectively~\cite{brandenburg:2019:arxiv}.
MPCP tries to benefit from data locality by distinguishing between local resources only used on one core and global resources.
Consequently, local resources reside in core-local memory, while global resources are placed in global memory.
DPCP takes a different approach to exploiting data locality by utilizing special synchronization processors, on which global resources are executed.
While a thread is executing on a synchronization processor, its host processor is free to use for other treads.
In its most basic version, all global critical sections are assigned to a single synchronization processor and the synchronization processor only executes critical sections.
In the generalized version of the protocol, multiple synchronization processors are allowed. Additionally, non-critical sections may be executed on synchronization processors (but will be preempted by critical sections).

%
%

We argue that contemporary and future processors call for a solution somewhere between MPCP and DPCP.
MBS can therefore be understood as the natural extension under the assumption of the availability of many cores.
The abundance of cores facilitates the exploitation of data locality in three ways~(see Section~\ref{sec:implementation}).
First, it allows the usage of many synchronization cores, ideally one per shared resource.
Second, synchronization cores can also be exclusively used for critical sections.
As a third consequence, there are no core-local resources, all resources are made global.

Oyama et al.~\cite{oyama:1999:pdsia} propose a similar form of control-flow migration where concurrent accesses to a resource are executed on a single core, the \emph{owner} or \emph{lock holder}.
The owner can, however, change over the course of execution, yielding no guaranteed cache locality.

A similar form of synchronization mechanism closely related to control-flow migration is \emph{delegation-based} synchronization, where some form of centralized server is responsible for the execution of the critical section.
As the execution of the critical section is handed over, these approaches can be viewed as a form of partial control-flow migration.
Examples include remote-core locking~\cite{lozi:2012:atc} and ffwd~\cite{roghanchi:2017:sosp}, as well as approaches centered on data structures like Flat combining~\cite{hendler:2010:spaa} and Actors~\cite{agha:1990:cacm}.

The concept of data locality is utilized in a similar fashion by a scheduling algorithm proposed by Boyd-Wickizer et al.~\cite{boyd:2009:hotos}, which revolves around bringing threads to the data they need to operate on.
The respective data objects are identified via source-code annotations.
MBS, on the other hand, is transparent for the application as it simply replaces existing locks.
Additionally, their approach does not consider timeliness constraints.

If timeliness is important, response-time analysis is the usual tool to guarantee timely execution of all threads.
In the presence of shared resources, response time is a combination of the thread's execution time, preemption by higher priority threads, and blocking time spent waiting for shared resources to be available~(that is, blocking time due to priority inversion). An example of such a formula can be found in~\cite{brandenburg:13:rtas}, in this case for partitioned, fixed-priority scheduling:

\begin{align}
  r_i &= e_i + b^l_i + b^r_i + \sum_{P(T_h) = P(T_i) \wedge h < i}{\lceil\frac{r_i + b^r_h}{p_h}\rceil\cdot e_h}
\end{align}

It consists of the following elements:
\begin{itemize}
  \item $r_i$: response time of task $i$
  \item $e_i$: worst-case execution time of task $i$
  \item $p_i$: period of task $i$
  \item $b^l_i$: priority-inversion blocking time caused by a task on the same core
  \item $b^r_i$: priority-inversion blocking time caused by a task on a remote core
  \item $P(T_i)$: current processor of $T_i$
\end{itemize}

The blocking-time parts of the formula~($b^l_i$, $b^r_i$) are highly dependent on the maximum number of times a thread can have to wait and can be derived by blocking-bound analysis~\cite{brandenburg:13:rtas,biondi:2016:rtss,yang:2015:rtss}, which yields upper bounds on the maximum a thread can be blocked due to waiting for a resource.
The complexity of such analyses depends both on the used synchronization protocol and the resource access structure of the application.
If critical sections are nested, the problem is NP-hard~\cite{wieder:2014:rtss}.

Detrimental to all those timing analyses are the effects of memory-related interferences, mostly through caches, as the timing difference between a cache miss and a cache hit has a significant impact on the timing behavior of a program.
The ability to predict such accesses correctly thus determines the accuracy of timing analyses.
Especially data caches are, however, highly unpredictable~\cite{lundqvist:99:rtcsa}, which gave rise to a wide field of research~\cite{lv:2016:lites} still actively worked on today~\cite{stock:2019:rtss}.
The problem is aggravated through inter-core interferences in multi-core shared-memory environments~\cite{wegener:2017:wcet}.
More complex hardware features in modern processors further hamper predictability~\cite{kaestner:19:wcet}.
Hybrid measurement-based approaches can overcome this problem, the quality of results is, however, highly dependent on the measurement data~\cite{kaestner:19:wcet}.
The guaranteed data locality of critical sections in MBS, therefore, facilitates timing analyses and helps to derive tight bounds on worst-case execution time~(WCET) and worst-case response time~(WCRT).

A different kind of memory-related overheads in the context of migration are the direct costs associated with the transfer of data.
The magnitude of these migration costs depends on the specific hardware platform and memory hierarchy and  may or may not be significantly large~\cite{bastoni:10:ospert,calandrino:06:rtss}.
In any case, these costs can be mitigated, for example, through hardware-based mechanisms~\cite{sarkar:09:sigplan}.
Such approaches can be combined with MBS to increase data locality and predictability with regard to thread-related data.


%% file: sec/implementation.tex
\section{Design and Implementation}
\label{sec:implementation}

\begin{figure}
 \includegraphics[page=1]{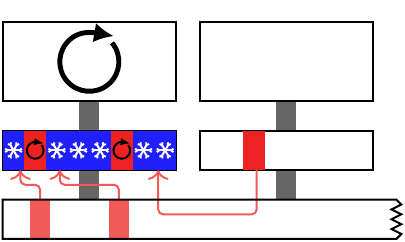}%
 \hfill
 \includegraphics[page=2]{tikz/lock.pdf}%

 \vspace{.2cm}
 \includegraphics[page=3]{tikz/lock.pdf}%
 \hfill
 \includegraphics[page=4]{tikz/lock.pdf}%
 \caption{Cache effects of locks (top) and MBS (bottom) when entering (left) and leaving (right) a critical section. Locks lead to poor data locality because shared data is not yet in the cache when entering and forbidden to access when leaving a critical section. MBS, in contrast, migrates the control flow to the known data location, keeping caches hot.}
 \label{fig:overview}
\end{figure}

This section discusses the design and implementation of MBS.
MBS operates under several assumptions that target real-time systems.
First, the set of locks is statically known, which is needed for static analysis, and also for the selection of synchronization cores.
Second, the system uses coarse-grained locking to protect its data structures from concurrent accesses.

\subsection{Synchronization by Migration}

Central to the concept of MBS is to consider a processor core as a resource that is shared between threads, but the scheduler assigns this resource to at most one thread at any moment in time.
MBS exploits this exclusive assignment for thread synchronization, and considers a processor core a synchronization object (i.e., a ``synchronization core'').
A synchronization core is a resource that is granted sequentially to threads that request this resource, and there is no forced withdrawal of this resource.
The acquisition of this resource is the migration of the current control flow to this core, and to release it, the control flow migrates back.
This migration can be implemented at library or operating system level and hidden behind a traditional \texttt{lock()}/\texttt{unlock()} interface.

In addition to the provision of mutual exclusion, MBS affects the shared data protected by this synchronization scheme.
This data is only accessed from critical sections, which are always executed on the synchronization core.
In consequence, this data can be placed in the core-local memory (i.e., first-level caches or scratchpads) of the synchronization core, and the data can permanently remain there since no other core accesses it---and only control flows that access this data migrate to this core.
Thus, control flow is moved to data, rather than vice versa.
In consequence, data accesses in critical sections benefit from guaranteed data locality.

Figure~\ref{fig:overview} summarizes data locality of locks and MBS when entering or leaving a critical section.
When a lock-based system enters a critical section, the wanted shared data is likely not in its local cache. Hence, caches are cold. When leaving the critical section, the shared data is still in the cache, but access is forbidden. Again, the cache contains data that is not used in the near future.
With MBS, in contrast, shared data is at a well-known location and only a part of the thread-related data (i.e., thread control block and top of stack) needs to be copied between cores. In consequence, caches are mostly hot.

The principal concept of MBS is independent of the scheduling policy used for the application cores.
Whether using fixed or dynamic priorities, whether using partitioned or global scheduling, the cache locality of shared date on the synchronization cores can always be exploited.
Only the degree to which cache-locality on the application core after a critical section can be used in timing analyses depends on the scheduling policy, with partitioned fixed-priority scheduling being the easiest to predict.

\subsection{Synchronization Cores}

For MBS, a synchronization core is a processor core that constitutes an exclusive resource.
Access is granted sequentially by a core-local scheduler which considers the system-wide resource-acquisition policy (e.g., FIFO or priority-based).

To ensure mutual exclusion, synchronization cores operate without preemption and all critical sections have a run-to-completion semantics.
In particular, the synchronization core may not participate in any other migration scheme, such as load balancing.
However, these restrictions only apply to synchronization cores.
Other cores (i.e., ``application cores'') may still support preemption or load balancing.

While in a critical section, the original core of a thread becomes available.
Multiple options exist for this original core.
First, the original core may be granted to another ready thread.
However, this thread necessarily has a lower priority, which makes this option functionally different to a lock-based system.
Second, a thread may leave a priority-based reservation that prevents other threads from running.
We call this variant MBS+R.
Third, the core may enter a low-power sleep state to save power.
In particular in power-constrained systems, this strategy helps to maintain a global power budget.
However, this strategy can only be efficient if the caches do not lose their data when sleeping.

\subsection{Data Pinning}

Several options exist to place shared data in the local memory of synchronization cores, depending on the hardware.
First, data can be placed in a cache \emph{implicitly} by simply accessing it, and the first memory access loads the data into the cache.
However, with this implicit scheme, data may be evicted, in particular, if the working-set size of a thread in the critical section exceeds the cache's capacity.
Second, some hardware architectures support \emph{explicit} cache management via cache-line pinning.
Then, the data is loaded and pinned during system initialization, and no eviction can occur.
Only thread-related data may be evicted, but the shared data remains certainly cached.
Third, some hardware primarily targeting hard real-time systems provide core-local scratchpad memory, where the software controls the contents of the local memory, and the hardware does not provide cache coherence.
Like explicit cache management, the system can place the shared data in the local scratchpad memory during system initialization.
The lack of cache coherence is no problem with MBS since shared data is only accessed from the synchronization core.
In addition, implicit eviction can be avoided since the scratchpad is software-controlled.

\subsection{Overheads}

The overheads of MBS are two-fold.
First, there is a \emph{dynamic} overhead due to thread migration.
A similar lock-based system, however, would also have synchronization overheads, in particular when blocking a processor core leads loss of cache state; and the copy operations of shared data as shown in Figure~\ref{fig:overview}.
Second, there is a \emph{static} overhead of dedicating a processor core for synchronization.
This core and its local memory are no longer freely available to the application.
However, this overhead is counteracted by improved data locality, which leads to lower and more predictable execution times, whereas lock-based systems suffer from locality-related interferences and overly pessimistic analyses.
The detailed influence of MBS on the execution time is discussed in Sections~\ref{ssec:wcet} and~\ref{ssec:wcrt}.

In summary, MBS trades notoriously difficult to predict cache-related interferences for well-predictable control-flow migration.
The trade-off is that the thread, parts of its stack (i.e., thread-local data), and other potentially used global data, has to be migrated to the synchronization core for each critical section. The amount of data that needs to be transferred can be determined exactly via static analysis, as long as the migration points are known~\cite{klaus:19:ospert}. For MBS, all migration points are calls to \texttt{lock()} and \texttt{unlock()} functions.

\subsection{Effects on WCET Analysis}\label{ssec:wcet}

MBS has the potential to affect both WCET and WCRT analysis positively.
With MBS, there is no need to model cache interference between cores as all accesses to shared data are isolated to dedicated cores.
Bus interference remains but can be handled through other means such as timing arbitration~\cite{kelter:11:ecrts} or hardware support~\cite{farshchi:20:rtas}.
For the critical sections of threads, the cache analysis becomes much easier as the shared data already resides in the caches of the synchronization cores\footnote{In the extreme case where the shared data is too large to fit in the core-local memory, there exists at least the potential to improve cache analysis by having more knowledge about the state of the caches.}.
As a result, tighter WCET bounds for the critical sections can be obtained.
Whether these theoretically possible benefits can be found in practical WCET analysis, depends on the capabilities of the used tools and techniques.

In measurement-based timing analysis, the tighter execution times achieved by the use of MBS can be observed directly in the measurements.
The simple nature of this technique allows it to be employed in basically every system with its disadvantage of the uncertainty if the worst case was actually captured in the measurements.

More reliable results can be obtained with the use of hybrid analysis techniques which combine measured timing data with static analysis techniques~\cite{kaestner:19:wcet}.
As the timing behavior is determined via measurements, the reduced execution times that come with MBS again directly affect the analysis results yielding tighter WCET bounds.

For pure static analyses, the effect of MBS visible in the analysis results highly depends on the underlying hardware model the analysis tool provides for the platform.
Only if it captures the caching behavior and does not make overly pessimistic assumptions elsewhere~(for example, bus arbitration), the positive effects of MBS are reflected in the WCET bounds.

\subsection{Effects on Response-Time Analysis}\label{ssec:wcrt}

Response-Time analysis benefits in two ways from MBS: indirectly from the tighter WCET bounds, and directly through easier blocking bound analyses.
The execution time of critical sections is included many times in response-time calculations, as it contributes directly to thread WCETs~($e$) and, by that, also to the blocking times~($b^l$, $b^r$).
As MBS enables tighter WCET bounds due to the known cache state, bounds on the response time also become tighter.

This improvement is only achieved as a trade-off with the newly introduced control-flow migration overhead, as, under MBS, a certain amount of data~(necessary stack-local data, thread-control block) has to be transferred to the synchronization cores.
Similar effects occur in suspending locks as the thread state has to be stored to and restored from memory before and after the suspension.
In MBS, this store and restore simply happens in the memory of another core instead of the own core.
With regard to response time, the overhead introduced by migration can be precisely bounded, as the data to be transferred can be statically determined.
Response-time Analysis thus improves, as uncertain cache states are traded with known cache states and predictable migration overhead.

Blocking-bound analyses already exist for various synchronization mechanisms.
Schedules produced by the use of MBS are essentially equivalent to lock-based approaches.
But as critical sections execute non-preemptively, the behavior of MBS also resembles priority-ceiling approaches~\cite{sha:90:tc}.
The main difference is the possibility for other~(MBS) or at least higher priority~(MBS+R) threads to execute during the execution of a critical section.
From this point of view, MBS+R behaves like stack-based priority-ceiling protocols~\cite{baker:90:rtss}: During the execution of a critical section, only unrelated higher priority threads are allowed to run.
MBS differs as it enables the execution of the critical section and unrelated higher priority threads in parallel.
Therefore, MBS will never lead to worse response times than traditional locking mechanisms.
Due to these similarities, we hope to be able to reuse known analysis techniques for existing synchronization approaches.

One of the major problems in bounding the blocking time, unbounded priority inversion, is avoided by design in MBS, as the scheduler on the synchronization cores grant access to resources non-preemptively and in accordance with the priorities of the waiting threads.
Additionally, MBS effectively decouples processor assignment and resource assignment.
An independent thread with medium priority cannot interfere with the execution of critical sections of a lower priority thread and in turn negatively affect a high priority thread waiting on the resource the lower priority thread currently holds.
As a consequence, a thread can be blocked at most one time by either a higher or lower priority thread for each of its critical sections, which can be efficiently bounded by the longest critical-section execution time of all users of a resource.

\subsection{Lock Nesting}

Nested locks are considered difficult for timing analysis~\cite{wieder:2014:rtss}, primarily due to the effect of \emph{transitive} blocking, where a thread holding resource $R^A$ and waiting for another resource $R^B$ potentially delays other threads that want to acquire resource $R^A$. Thus, the contention for resource $R^B$ affects all threads working with $R^A$.
In addition to timing analysis, lock nesting is prone to deadlocks.
However, if a sound worst-case blocking-bound analysis yields a finite duration, deadlocks are not possible.
Solutions for lock nesting in real-time systems are either to forbid nesting entirely, to group locks that may nest to a single lock~\cite{brandenburg:2019:arxiv}, or to analyze transitive blocking~\cite{biondi:2016:rtss}.

Since MBS can result in the same schedule as using locks, existing approaches can be reused.
First, systems that disallow lock nesting and systems that replace nested locks with group locks are trivially supported.
Second, it is compatible with blocking-bound analysis techniques like~\cite{biondi:2016:rtss}, where locks are granted in FIFO order.
Third, priority-based approaches are supported by executing idle threads with elevated priorities.

\subsection{Integration in \zephyr}

For demonstration, we have integrated MBS in \zephyr~\cite{zephyr:web}, an embedded operating system for the Internet of Things (IoT) that has recently started to support multi-core processors.
We implement MBS in \zephyr on a Raspberry 3B+, a typical platform for IoT systems with soft real-time requirements.

Our integration of MBS into \zephyr is relatively straightforward.
In particular, the scheduler requires some adaptions for synchronization cores (in our case, only one core).
If a thread acquires an ``MBS mutex'' (which can coexist with ordinary mutexes), it is placed in a dedicated waiting queue for the corresponding synchronization core.
The scheduler on the synchronization core then dequeues the thread and executes it normally.
Migration back is implemented by placing the thread back in the ordinary ready list (i.e., the ready list for application cores).
A second implementation variant, MBS+R, leaves a reservation on the core by marking it as ``blocked due to a thread that is currently migrated''.
While this flag is set, the idle thread is scheduled, until the thread completes its critical section and migrates back.

In Zephyr, the scheduler and its data structures are protected by a global spin-lock with interrupt suppression.
This global lock cannot be implemented in MBS because it protects the scheduler which implements MBS.
Timing analysis of this global lock can therefore not benefit from MBS.
However, a scheduler lock is a common implementation technique for embedded multi-core systems.
System-wide timing analyses therefore either have to analyze its blocking time, or alternatively use wait-free techniques that are unaffected by blocking.

%% file: sec/evaluation.tex
\section{Evaluation}
\label{sec:evaluation}

\newcommand{\variant}[1]{\texttt{#1}\xspace}

This section evaluates our implementation of MBS in the context of a soft and hard real-time systems. We evaluate execution times \emph{experimentally} and we also apply static and hybrid timing analysis.

For the first part of the evaluation, we use a Raspberry Pi 3B+ running a modified Zephyr OS that supports MBS.
The processor has four cores with 32\,KiB of core-local L1 data cache each, and 512\,KiB of shared L2 cache.
We select one core as a dedicated synchronization core, and three application cores remain.
Cache pinning is implicit---data is loaded to the cache on first access and evicted automatically.
This strategy allows us to also evaluate scenarios where the shared data exceeds the L1 cache capacity, where eviction is inevitable.

We use a family of microbenchmarks that allow for precise control of the thread-local working set size and the shared data size.
These benchmarks access a local and a shared buffer, where the local buffer has size $\lambda$ and is not synchronized, and the shared buffer of size $\sigma$ requires mutual exclusion.
Each benchmark cycle iterates over both buffers alternatingly.
Elements in each buffer are modified sequentially, modifying each cache line once per benchmark cycle.

We compare the following synchronization variants. First, a \variant{MBS} version without original core reservations that uses four application threads. Second, a \variant{MBS+R} variant with original core reservation uses only three application threads. In addition, standard \zephyr \variant{spinlock} and \variant{mutex} implementations serve as a baseline, both with four threads.

In the second part, we use an Infineon TriCore TC397 as evaluation platform, which features a complex memory hierarchy.
It provides one level of core-local 16\,KiB caches for access to global memory as well as separate core-local scratchpad memories.
We use this as an exemplary prototype to examine the influence of MBS on worst-case analyses.

\subsection{Measurement-based Performance Evaluation}

\begin{figure}
 \includegraphics{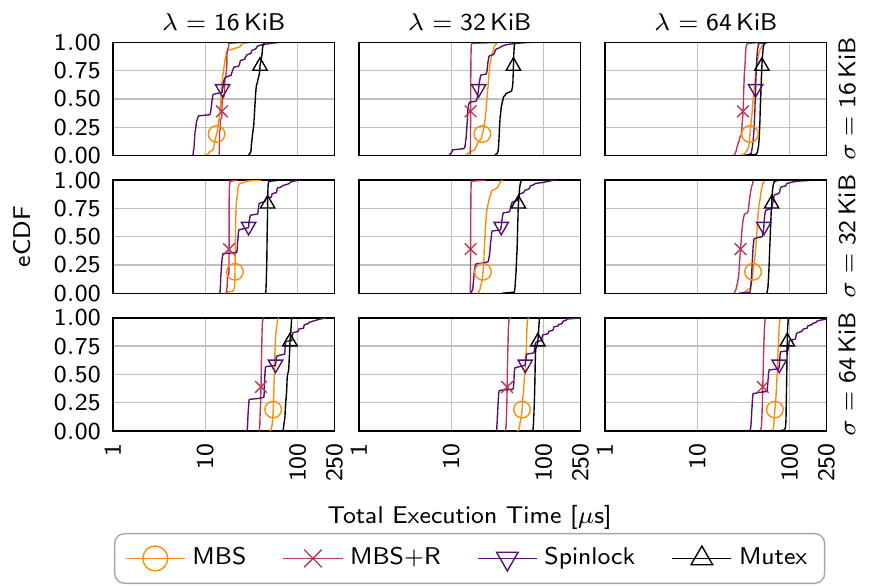}
 \caption{Latency distributions of critical \& non-critical sections combined.}
 \label{fig:latency}
\end{figure}

\begin{figure}
 \includegraphics{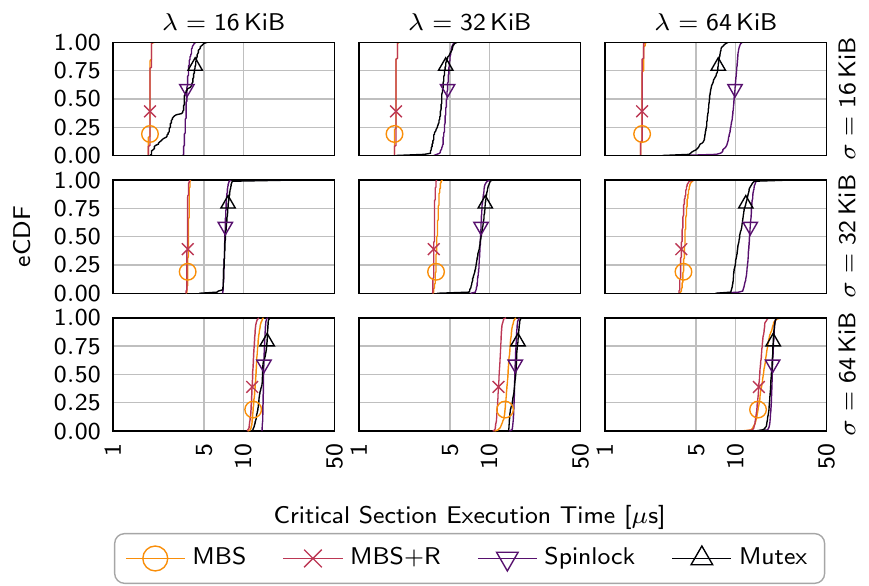}
 \caption{Latency distributions of critical section only.}
 \label{fig:crise}
\end{figure}

Figure~\ref{fig:latency} summarizes the latency of a full benchmark cycle for each microbenchmark version.
We vary the sizes $\lambda$ and $\sigma$ so that the buffer is either too small, too large, or roughly fitting the core-local L1 cache.
Further data structure sizes are omitted to maintain readability, but the corresponding experiments confirm the results discussed in the following.
Each sub-plot in Figure~\ref{fig:latency} visualizes the latency distribution for a combination of buffer sizes.
In total, the performance of MBS, locks, and mutexes is of similar magnitude.
However, MBS has benefits over locks and mutexes.
First, the MBS variants have a lower latency variation than locks and mutexes.
Second, the observed worst-case latency and 99\,\% tail latency are reduced by MBS, which is relevant for soft real-time systems.
Third, when the shared buffer fits the core-local L1 cache, MBS is faster than locks and mutexes.

Figure~\ref{fig:crise} visualizes the latency of only the critical sections in the same experiment as Figure~\ref{fig:latency}.
It shows clearly that, if the shared data fits the L1 cache, MBS is faster.
This result indicates that, as expected, the improved cache locality results in faster critical section execution time.
If, however, the shared data is too large for the L1 cache, MBS has little benefit since it cannot exploit the cache optimally.

\begin{figure}
 \includegraphics{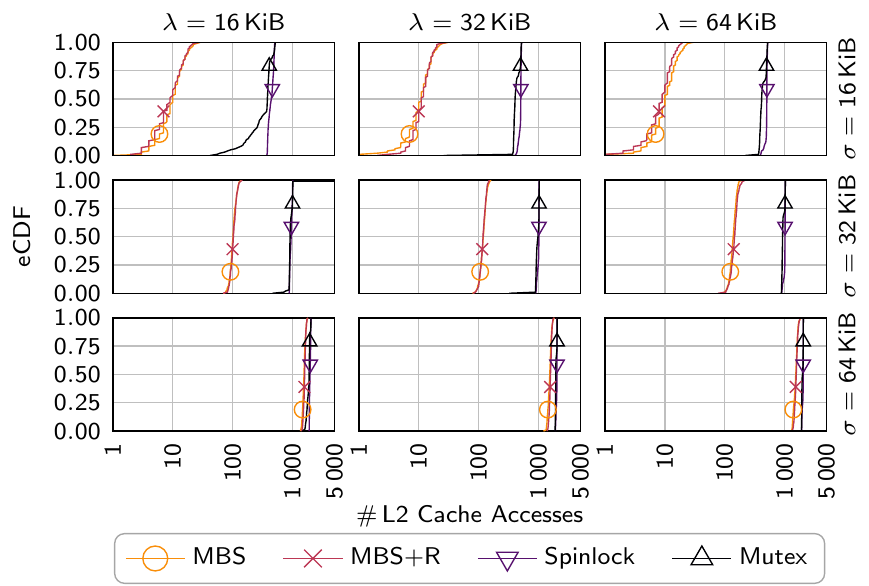}
 \caption{Reduction in shared L2 cache accesses as data remains in the core-local L1 cache.}
 \label{fig:cache}
\end{figure}

Figure~\ref{fig:cache} summarises the cache locality of our experiments.
It visualizes the number of L2 cache accesses inside critical sections.
Since this cache is shared, it is only accessed if the core-local L1 cache does not contain the requested data.
The graph shows that MBS significantly reduces the number of L2 accesses---most data accesses are handled by the L1 cache of the synchronization core.
Again, this requrires that the shared data fits into the L1 cache.
If the shared data is too large, the number of L1 misses is similar for all examined variants.

\subsection{WCET Analysis}

As described in Section~\ref{ssec:wcet}, MBS influences different timing analysis techniques differently.
Therefore, we conduct access-latency experiments to data in global and local memory on the TriCore platform and compare the WCET bounds of multiple analysis tools.
The size of the shared data either fits well in the cache, fits exactly in the cache or is too large for the cache.
Timing analysis is performed by tracing as an example of measurement-based approaches, TimeWeaver~\cite{kaestner:19:wcet} as a hybrid analysis tool, and aiT~\cite{ait}\footnote{Version: 21.04} as a purely static WCET tool.
We additionally use aiT in a variant, where all accesses are assumed to result in cache hits.
While this is not a valid WCET analysis technique, it allows us insight into the improvements possible with the known cache locality of MBS.

\begin{figure}
  \centering
  \includegraphics{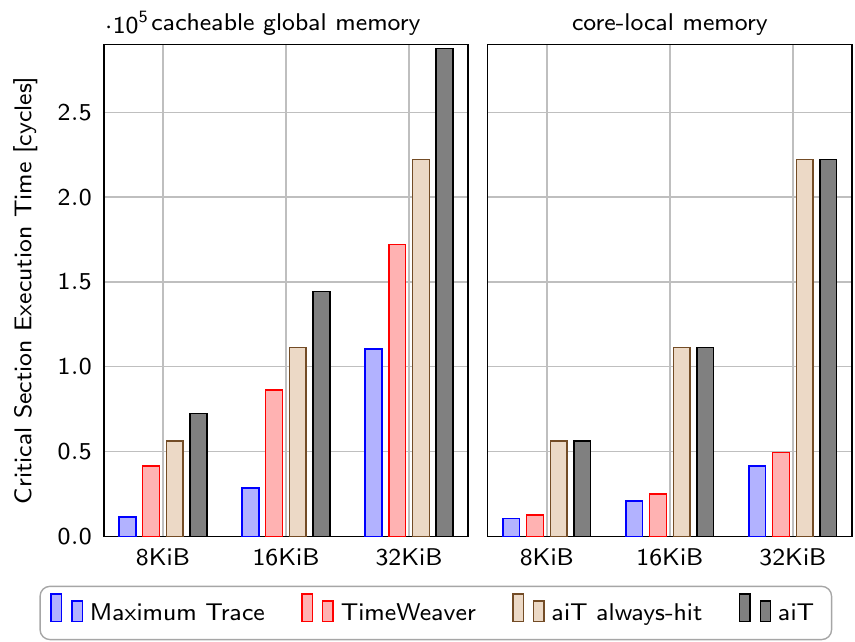}
  \caption{WCET estimates for accesses to globally shared data and core-local data with different analysis techniques.}
  \label{fig:wcet}
\end{figure}

With hardware tracing, we can directly observe the effects of data locality in the measurement results.
Figure~\ref{fig:wcet} shows the expected increase in execution time once the shared data does not fit completely into the cache~(32\,KiB).
Using the larger core-local memory, no such increase occurs~(see Figure~\ref{fig:wcet}). Additionally, the access times are generally shorter.
These results showcase the potential which lies in the optimal utilization of the memory hierarchy for access to shared data.
With measurement-based timing analysis, we can benefit from tighter bounds on execution time by employing MBS.

Similar results arise with the use of the TimeWeaver tool.
The WCET bounds follow the trend of the measurement data in Figure~\ref{fig:wcet}, although overestimations exist in the cache scenario.
In the case of purely core-local accesses, however, the bounds are much closer to the measured data.
Hybrid analysis techniques can, thus, also profit from MBS.
To which degree depends on the model of the hardware platform.

This brings us to the bounds obtained by static analysis based on sophisticated hardware models.
In Figure~\ref{fig:wcet}, we can also see the differences between the aiT results with (pessimistic) normal and always-hit cache behavior, affirming the potential improvements by data locality.
Unsurprisingly, both variants yield the same results in Figure~\ref{fig:wcet}, as no caches are involved in the data accesses.
The results are, however, pessimistic overestimations, which we cannot explain through the memory hierarchy.
It is possible that the behavior of the core-local accesses is correctly captured, but the tight access bounds are overshadowed by pessimistic assumptions in other parts of the analysis.
These results underline that the accuracy of the hardware model highly influences how strongly the positive effects of MBS are visible.

\subsection{Discussion}

The experiments show that, performance-wise, the costs of locking and the costs of thread migration are similar.
However, soft real-time systems can utilize MBS to achieve a more constant performance.
Hard real-time systems, in comparison, rely on static analysis tools that have to support the knowledge on data locality.

%% file: sec/conclusion.tex
\section{Conclusion}
\label{sec:conclusion}

This paper has presented Migration-Based Synchronization (MBS), a sweet spot in multi-core real-time system design that simplifies synchronization, cache analyses, and blocking-bound analyses, for both hard and soft real-time systems.
In particular, MBS achieves a system design where shared data has statically known locations in the memory hierarchy.
Besides analysis simplification, MBS also reduces the execution time (in particular, for the worst case) by improving data locality.
Thus, hard real-time systems can benefit from a simplified and improved cache-locality analysis, reduced critical section WCET, and in consequence, reduced blocking bounds.
Soft real-time systems also benefit from reduced cache-related interferences and lower latency variation.

Our evaluation confirms the reduction of (core-local) L1 cache misses, which reduces latency variation.
MBS mainly benefits if the shared data fits in the core-local cache of the synchronization core.
Both soft and hard real-time systems benefit from known cache hits.

When considering future many-core real-time systems, the cost of MBS scales better than traditional approaches.
The main cost (i.e., a dedicated synchronization core) becomes increasingly acceptable, considering that the benefits are improved analysability and reduced latencies.
In comparison, existing approaches have to pessimistically assume increased cache-related interferences and transitive blocking times.

Future work will further combine MBS with WCET analysis tools for hard real-time systems.
We intend to provide tools that map resources to synchronization cores, optimally selecting either locks or MBS.
This trade-off is necessary because current-generation hardware still has a relatively small number of cores.
In addition, the migration costs, the capacity of core-local caches, and data structure sizes need consideration.
Another approach is the run-time transition between MBS and locks, for example, in mixed-criticality systems.
In such a scenario, a processor core may be used for arbitrary application-specific workload in the normal case but used as a synchronization core if it is needed.
These optimizations are only possible because MBS does not require code-level application changes---instead, migration can be hidden behind a traditional \texttt{lock()}/\texttt{unlock()} interface.